\documentclass[%
 reprint,
 amsmath,amssymb,
 aps,
]{revtex4-2}

\usepackage{graphicx}% Include figure files
\usepackage{dcolumn}% Align table columns on decimal point
\usepackage{bm}% bold math
\begin{document}

%\preprint{APS/123-QED}

\title{Benchmarking simulation of hybrid decoding scheme for parity-encoded spin systems}

\author{Yoshihiro Nambu}
\email{y-nambu@aist.go.jp}
\affiliation{NEC-AIST Quantum Technology Cooperative Research Laboratory~~\\
 National Institute of Advanced Industrial Science and Technology }

\date{\today}
\begin{abstract}
This paper presents classical benchmark simulations of a practical hybrid decoding scheme for parity-encoded spin systems, which is well-suited to the development of quantum annealing devices based on on-chip superconducting technology. We compared the performance of finding the optimal solution using two embedding schemes for emulating all-to-all connectivity from local interactions: the SLHZ scheme, proposed by Sourlas, Lechner, Hauke, and Zoller, and the commonly used minor embedding (ME) scheme. We found that the SLHZ scheme is more efficient than the ME scheme when combined with postreadout classical decoding based on the classical bit-flipping algorithm, although the SLHZ scheme itself is substantially less efficient than the ME scheme. 
\end{abstract}
\maketitle
Quantum annealing (QA) is expected to hold great potential as a fast solver for the hard combinatorial optimization problem \cite{kadowakiQuantumAnnealingTransverse1998,RevModPhys.90.015002}. It is well-known that most problems, such as a quadratic unconstrained binary optimization problem (QUBO), can be mapped to 2-local couplings between spin pairs \cite{568530b2-62d5-3d43-9e2f-adfdf424006b,kirkpatrickOptimizationSimulatedAnnealing1983,10.3389/fphy.2014.00005}. Implementing a large number of spins with controllable all-to-all pairwise connectivity is required for QA devices. However, this demand is challenging in actual devices because it requires individually programmable long-range interactions. In practice, this is incompatible with interactions that are implemented via short-range physical links in actual devices. So, several techniques can be used to circumvent this problem. One solution is based on the minor embedding (ME) scheme, named from graph theory \cite{choiMinorembeddingAdiabaticQuantum2008a,choiMinorembeddingAdiabaticQuantum2011a}. In the ME scheme, the $N$ spins given by the original Ising problem, which we refer to as the logical spins, are replaced by chains of physical spins implemented on the device. Strong ferromagnetic interactions within an individual chain impose energy penalties to align the physical spins to behave as a single logical spin. As a result, the $N$ spins of the logical problem are encoded into $K=\mathcal{O}(N^2)$ physical spins. The ME scheme is actually used in D-Wave QA devices \cite{pudenzErrorcorrectedQuantumAnnealing2014,pudenzQuantumAnnealingCorrection2015,vinciQuantumAnnealingCorrection2015}. 

Alternatively, Lechner, Hauke, and Zoller have proposed another clever embedding scheme \cite{lechnerQuantumAnnealingArchitecture2015}. In their scheme, $K=\binom{N}{2}$ physical spins are arranged on a two-dimensional lattice with a square unit cell, and the four spins at each corner of the lattice are coupled via a local four-body interaction. The $K$ physical spins encode the parity, i.e., whether they are aligned or anti-aligned, of $K$ possible pairs of logical spins, and the $K$ logical couplings are mapped to local fields acting on the $K$ physical spins, respectively. It is interesting to note that the same scheme was previously proposed by Sourlas as an instance of the soft-annealing concept based on an isomorphism between the Ising model and the classical error-correcting codes (ECC) \cite{sourlasSpinglassModelsErrorcorrecting1989,sourlasSpinGlassesErrorCorrecting1994,sourlasSoftAnnealingNew2005}.  From the viewpoint of the ECC, the values of physical spins correspond to the codeword, and the product of the values of the four physical spins at the corners of the unit cell corresponds to the syndrome. The four-body interactions within the unit cell impose an energy penalty on each syndrome to satisfy the parity constraints of ECC. This scheme will be referred to as the SLHZ scheme hereafter. The advantages of this scheme are as follows. First, because it enables a quasiplanar layout, it is well-suited to developing QA devices based on on-chip superconducting technology\cite{PhysRevLett.53.1260,PhysRevLett.55.1543,PhysRevLett.55.1908,nakamuraCoherentControlMacroscopic1999}. Actually, several researchers are conducting studies based on the SLHZ scheme and superconducting device technology\cite{puriQuantumAnnealingAlltoall2017a,niggRobustQuantumOptimizer2017a,zhaoTwoPhotonDrivenKerr2018,onoderaQuantumAnnealerFully2020,yamajiCorrelatedOscillationsKerr2023}. Second, this scheme demonstrates excellent scalability in hardware implementation when the number of logical qubits increases. Third, since the SLHZ scheme is closely related to the low-density parity-check (LDPC) code  \cite{pastawskiErrorCorrectionEncoded2016}, we can use classical decoding techniques to speed up the search for an optimal solution, as shown below. 

Both these schemes require the same $K=\mathcal{O}(N^2)$ physical resource cost when the logical $N$ spins are embedded into them. These costs are obviously higher than the cost of the original logical problem. As a result, the penalty occurs for both schemes. In this paper, we present the benchmarking experiment for these embedding schemes and show that the overhead depends not only on the number of physical spins but also on other important factors. We suggest that although the overhead of the original SLHZ scheme is far larger than that of the ME scheme \cite{albashSimulatedquantumannealingComparisonAlltoall2016}, we can cancel the overhead by classical postreadout decoding. We believe that this suggests the potential of the SLHZ scheme over the ME scheme as a platform for the superconducting QA devices.  

We used classical Markov Chain Monte Carlo (MCMC) sampler to benchmark the different embedding schemes. Although this benchmark is not necessarily reflect the performance of the actual QA devices, it is enough to gain insight about what is important to reduce the overhead in the SLHZ scheme. We generated temporal sequence by using MCMC sampler. The ensemble of the samples in the sequence have statistical distribution associated with the Gibbs ensemble. We investigated the probability that the ensemble, as generated above or the randomly generated ensemble explained below, involve the optimal solution corresponding to the logical ground state, which will be referred to as the success probability, hereafter. We evaluated the success probabilities as a function of sample size and compared the performance of each scheme based on the results. If the same probability is obtained with fewer samples, or if a higher probability is obtained with the same number of samples, we decide the performance is superior.

\begin{figure*}
\includegraphics[viewport=160bp 140bp 800bp 450bp,clip,scale=0.8]{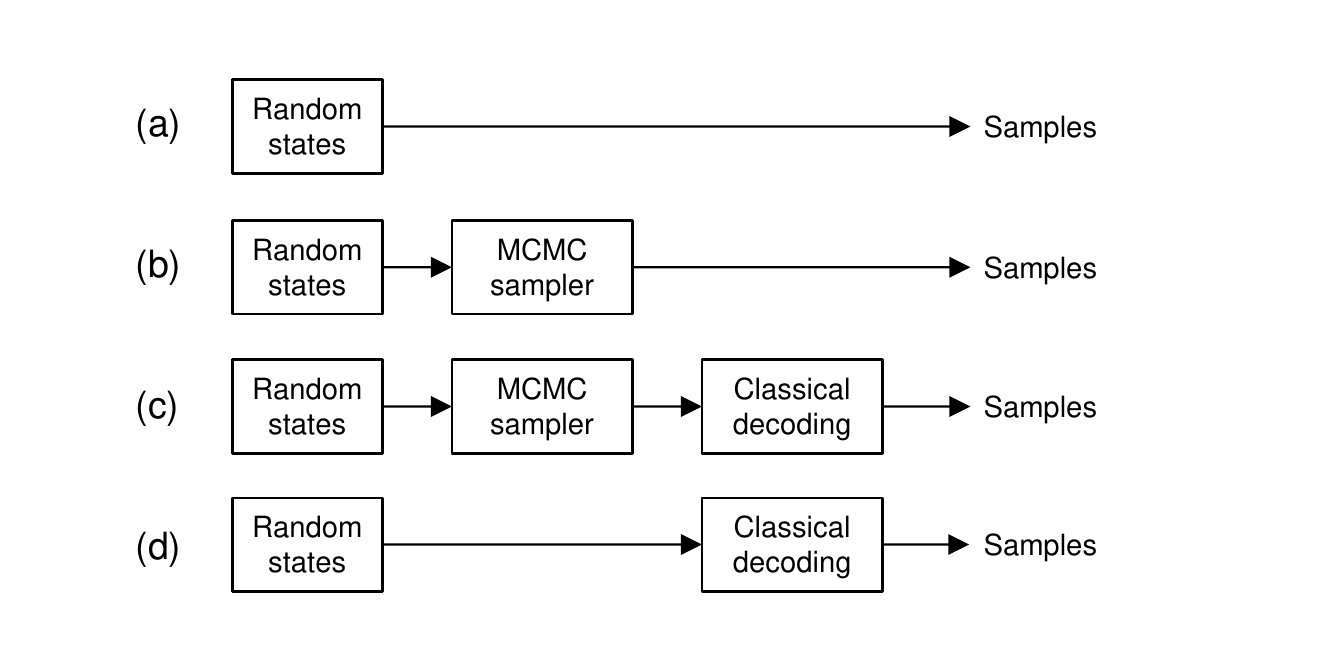}
\caption{Conducted four types of benchmarking experiments. These correspond to: (a) exhaustive search of the optimal state, (b) search based on the classical annealing, (c) search based on the classical annealing combined with postreadout classical decoding, and (d) exhaustive search combined with postreadout classical decoding.}
\label{fig:1}
\end{figure*}

In this work, four benchmark experiments schematically shown in Fig.\ref{fig:1} (a)-(d) were conducted for every embedding scheme presented later: 
\begin{enumerate}
\item[(a)] As theoretical baseline data, we evaluated the success probability using the $M$ randomly generated samples, which amounts to exhaustive search of the optimal state. 
\item[(b)] We generated a temporal sequence of $M$ samples from a randomly generated sample using MCMC sampler, and evaluated the success probability using these $M$ samples. This experiment amounts to a classical annealing-based search.
\item[(c)] We made postreadout decoding for $M$ samples obtained by MCMC sampler, and evaluated the success probability using the decoded samples. This experiment amounts to a classical annealing-based search combined with classical postreadout decoding.
\item[(d)] Additionally, we made postreadout decoding for $M$ randomly generated samples, and evaluated the success probability using the decoded samples. This experiment amounts to exhaustive search combined with classical postreadout decoding.
\end{enumerate}
\begin{figure}[tb]
\includegraphics[viewport=360bp 200bp 620bp 400bp,clip,scale=0.8]{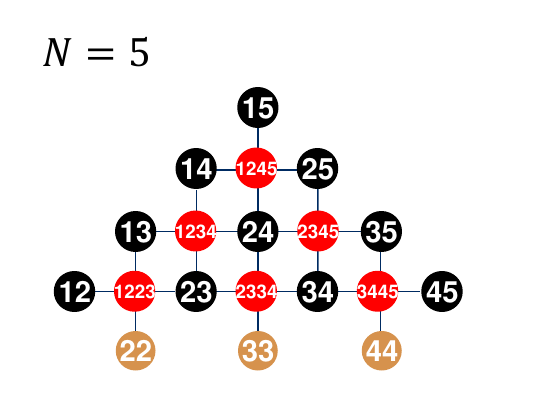}
\caption{Architecture of the SLHZ scheme for $N=5$ logical spins. The black and red circles indicate physical spins and syndromes, respectively. Numbers within the circles define the indices for each spin and syndrome. }
\label{fig:2}
\end{figure}
We considered the optimization based on the following all-to-all connected logical Ising Hamiltonian: 
\begin{equation}
H^{logical}(\textbf{Z})=-\sum_{i<j }J_{ij}Z_{i}Z_{j},
\label{eq:1}
\end{equation}
where $Z_{i}\in \{\pm1\}$ and $\textbf{Z}=\{Z_{i}\}\in\{\pm 1\}^{N}$ represent variable of the logical spin $i\in[1,N]$ and the configuration of logical spins, respectively. The set of pairwise couplers $\textbf{J}=\{J_{ij}\}$ for any possible $\{i,j\}$ pairs with $i,j\in[1,N]$ encodes an optimization problem we wish to solve by finding the lowest energy spin configuration. Here, we absorbed the linear Ising terms into pairwise interaction terms as couplers to a fixed auxiliary spin, without loss of generality. We investigated the performance of the following three embedding schemes (i)-(iii) associated with this Hamiltonian.

\paragraph*{(i) SLHZ scheme}
Hamiltonian of the SLHZ scheme is given by \cite{lechnerQuantumAnnealingArchitecture2015} 
\begin{equation}
H^{SLHZ}(\textbf{z})=-\sum_{i<j}J_{ij}z_{ij}+C_{4}\sum_{\{ i,j,k,l\}\in P_{4}}\frac{1-s_{ijkl}^{(4)}(\textbf{z})}{2},
\label{eq:2}
\end{equation}
where $C_{4}$ is the positive real parameter to be adjusted, $z_{ij}\in \{\pm1\}$ and $\textbf{z}=\{z_{ij}\}\in\{\pm 1\}^{K}$ represent the variable of the physical spin $ij$ and the configuration of the physical spins, respectively. Figure \ref{fig:2} illustrates examples of the architecture of the SLHZ scheme, which also indicates the indexing of the physical spins. Based on this indexing, $s_{ijkl}^{(4)}$ is given by 
\begin{equation}
s_{ijkl}^{(4)}(\textbf{z})=z_{ik}z_{jk}z_{jl}z_{il}
\label{eq:3}
\end{equation}
with $z_{ii}=1$ fixed, which defines the weight-4 syndrome for parity encoding: $z_{ij}=Z_{i}Z_{j}$  for any possible pairs of $\{i,j\}$ with $i,j\in[1,N]$. The sum in the second term of the Hamiltonian (\ref{eq:2}) runs over the elements of the set $P_{4}$  of unit cells defined by a two-dimensional array of physical spins. Physically, the first term of the Hamiltonian (\ref{eq:2}) represents interaction between local fields and physical spins, and the second term represents four-body interactions among every four adjacent physical spins. On the other hand, if we interpret this Hamiltonian as the cost function for ECC, the first term represents correlation between the observed signal $\textbf{J}$ and $\textbf{z}$, and the second term represents parity constraints for $\textbf{z}$ \cite{sourlasSpinGlassesErrorCorrecting1994,nambuPracticalHybridDecoding2026}. The second term is minimized if and only if $\textbf{z}$ is codeword for the parity encoding, i.e., $z_{ij}=Z_{i}Z_{j}$ $ ^{\forall} \{i,j\}$. 

We need to search the solution $\textbf{z}^{*}=\{z^{*}_{ij}\}\in\{\pm 1\}^{K}$  given by 
\begin{equation}
\textbf{z}^{*}=\underset{\textbf{z}\in D}{\arg\min}H^{SLHZ}(\textbf{z}),
\label{eq:4}
\end{equation} 
where $D$ denotes the set of all codewords satisfying the parity constraints that minimizes the second term of the Hamiltonian (\ref{eq:2}). In this work, we search $\textbf{z}^{*}$ by using an MCMC sampler, which needs to optimize the parameter $C_{4}$. 

\paragraph*{(ii) Modified SLHZ scheme}
We consider modification of the Hamiltonian (\ref{eq:2}). According to general knowledge of ECC, parity constraints are not unique and can be enforced in various ways. For example, they may be chosen to be weight-4 as well as weight-3 parity constraints \cite{pastawskiErrorCorrectionEncoded2016}. If weight-4 parity constraints are allowed, we can imagine the associated spin system arranged as a two-dimensional array with geometrically local interactions, as shown above. We will discuss the modified scheme based on the weight-3 parity constraints below. The Hamiltonian is given by 
\begin{equation}
H^{SLHZ'}(\textbf{z})=-\sum_{\{ i,j\} }J_{ij}z_{ij}+C_{3}\sum_{\{ i,j,k\}\in P_{3}}\frac{1-s_{ijk}^{(3)}(\textbf{z})}{2},
\label{eq:5}
\end{equation} 
where
\begin{equation}
s_{ijk}^{(3)}(\textbf{z})=z_{ij}z_{jk}z_{ik}
\label{eq:6}
\end{equation}
is the weight-3 syndrome associated with the parity-encoding, and $P_{3}$ is the set of possible triplets $\{i,j,k\}$  with $i,j,k\in[1,N]$. This is quite natural and most simple choice from the viewpoint of the ECC. This choice treats all variables symmetrically but involves long-range couplers. Therefore, this scheme is unrealistic for implementing the actual QA device. Nevertheless, we are interested in this scheme. This is because comparing the SLHZ and modified SLHZ schemes reveals overhead of the SLHZ scheme, which must be borne when implementing the QA devices, avoiding long-range couplers. We need to search the solution $\textbf{z}^{*}$ similar to the Eq. (\ref{eq:4}) by optimizing the positive real parameter $C_{3}$.

For the two schemes above, we applied the same classical postreadout decoding algorithm in the benchmarking experiment (c). If we obtain readout $\textbf{r}=\textbf{z}^{*}\circ\textbf{e}=\{r_{ij}\}\in\{\pm 1\}^{K}$ with some errors $\textbf{e}\in\{\pm 1\}^{K}$ from the sampler, where $\circ$ denotes component-wise multiplication, we have a chance to recover $\textbf{z}^{*}$ by using postreadout decoding. It has been shown in Ref. \cite{nambuPracticalHybridDecoding2026} that the bit-flip (BF) algorithm provides a simple and practical classical postreadout decoding procedure for the SLHZ scheme. If we rearrange the elements of $\textbf{z}$ and $\textbf{r}$ to form symmetric matrices $\hat{\textbf{z}}$ and $\hat{\textbf{r}}$ whose elements are given by $z_{ij}(=z_{ji})$ and $r_{ij}(=r_{ji})$, respectively, the algorithm can be represented by iterative operation 
\begin{equation}
\hat{\textbf{z}}=\mathcal{F}^{(n)}(\hat{\textbf{r}})
\label{eq:7}
\end{equation}
with 
\begin{equation}
\mathcal{F}(\hat{\textbf{r}})=\mathrm{sign}\left[\hat{\textbf{r}}(\hat{\textbf{r}}-\textbf{I}_{K\times K})\right],
\label{eq:8}
\end{equation}
where $\textbf{I}_{K\times K}$ is the identity matrix, the sign function is component-wise, and the superscript $(n)$ for $\mathcal{F}$ indicates $n$ iterative operation. Note that $\hat{\textbf{z}}=\textbf{Z}^{T} \textbf{Z}$ if the logical state $\textbf{Z}=\{Z_{i}\}$ is represented as column vector. The matrix function  $\mathcal{F}$  is composed of iterations of matrix subtraction, product, and component-wise operation \cite{nambuPracticalHybridDecoding2026}. This decoding algorithm was derived based on weight-3 syndrome $s_{ijk}^{(3)}(\textbf{z})$, for which the symmetry about the permutation of the elements of $\textbf{z}$ was utilized. This algorithm can be viewed as an iterative algorithm that gradually increases the number of weight-3 syndromes with unit values by flipping spins according to the majority vote of the syndromes. As a result, it maps every readout $\textbf{r}$ to the codeword $\textbf{z}$ nearest to $\textbf{r}$ after $n=n_{0}$ iterations. If $\textbf{r}$ is near to $\textbf{z}^{*}$, we can successfully recover $\textbf{z}^{*}$. Experimental results suggest $n_{0}\sim 5$ or $6$  is enough independently of $K$. An iterative algorithm is essential because logical spins are redundantly and nonlocally encoded in the physical spins. As shown later, the function of this algorithm is confirmed by a benchmarking experiment (d).

\paragraph*{(iii) Minor embedding scheme}
We also made benchmarking of the ME scheme to compare its performance with those of the SLHZ and modified SLHZ schemes. In this work, we considered "Chimera" graph architecture discussed in Ref. \cite{albashSimulatedquantumannealingComparisonAlltoall2016} , which consists of the tiling of an $L \times L$ grid of $K_{4,4}$ cells. If we need to embed $N$ logical spins, we require a chain of $\left \lceil \frac{N}{4} \right \rceil+1$ physical spins per logical spin, and therefore require a total of $K=N\left(\left \lceil \frac{N}{4} \right \rceil+1 \right)$ physical spins. The associated Hamiltonian is formally written as
\begin{equation}
H^{ME}(\textbf{z}) = H^{ME}_{problem}(\textbf{J},\textbf{z})+C_{ME} H^{ME}_{chain}(\textbf{z}),
\label{eq:9}
\end{equation}
where the first term describes problem couplings that depends on $\textbf{J}$, while the second term describes chain couplings that is independent of $\textbf{J}$. The problem coupling term involves only short-range coupling within each $K_{4,4}$ cell, while the chain coupling term involves long-range couplings to interconnect $K_{4,4}$ cells and lowers the energy of chains of the physical spins when they are aligned. Since all the interactions are pairwise in this scheme, both terms can be written as a quadratic form. Then, both terms are characterized by $K \times K$ sparse and symmetric matrices.

To map the logical ground state of the Hamiltonian (\ref{eq:1}) to the physical ground state of the Hamiltonian (\ref{eq:7}),  we need to adjust the positive real parameter $C_{ME}$  and search the solution 
\begin{equation}
\textbf{z}^{*}=\underset{\textbf{z}\in D'}{\arg\min}H^{ME}(\textbf{z}),
\label{eq:10}
\end{equation} 
where $D'$ represents the set of all spin configurations in which the respective ferromagnetic chains described by $H^{ME}_{chain}(\textbf{x})$ are aligned. If some ferromagnetic chains are broken (where not all members of a chain are aligned), the spin configuration do not correspond to logical states and must be corrected. Majority vote (MV) is known for the most simple deterministic algorithm for postreadout error correction of ME scheme. In this method, the value of the logical spin is determined by the MV of the physical spin readout in the relevant chain. If it is tie, we determine it by coin-tossing. For a more complete exposition of the ME scheme, see, e.g., Ref. \cite{albashSimulatedquantumannealingComparisonAlltoall2016}.

We made benchmarking experiments (a)-(d) using a spin-glass problem for $N=14$  ($K_{14}$)  with couplings $J_{ij}\in[-\tfrac{1}{4},\tfrac{1}{4}]$ chosen uniformly at random as an example, which has been previously studied in Ref. \cite{nambuPracticalHybridDecoding2026}. The experiments have been performed using the \textit{Mathematica}$^{\circledR}$ Ver. 14 platform on the Windows 11 operating system. We have used rejection-free MCMC sampling, in which all self-loop transitions are removed from the standard MCMC \cite{nambuRejectionFreeMonteCarlo2022}. Starting from a random initial state, we sampled $\textbf{r}$ using a rejection-free MCMC sampler, and by running the MCMC loop $M$ times, we stored the sequence $\{ \textbf{r}\}$ with size $M$. The performance has been evaluated by repeating the independent and identical simulation 1000 times, yielding a set of 1000 sampled sequences $\{ \textbf{r}\}$. We have evaluated the average probability that $\{ \textbf{r}\}$ involves the optimal state $\textbf{z}^{*}$, which gives the success probability for the experiment (b). In the experiment (c), we have applied the appropriate postreadout decoding algorithm (the BF algorithm for the SLHZ and modified SLHZ schemes and the MV algorithm for the ME scheme) to the same sequence $\{ \textbf{r}\}$ obtained in the experiment (a) to obtain the decoded sequence $\{ \textbf{r}'\}$. We evaluated the success probability from a thousand decoded sequences $\{ \textbf{r}'\}$ in a similar way. Note that the results of the exhaustive search experiment (a) are trivial. If the logical ground state is not degenerated, the success probability for given $M$ samples is given by $P_{success}(M)=1-(1-p)^{M}$, where $p$ is given by $p=2^{-N}$ for the original logical $N$-spin problem, and $p=2^{-K}$ for embedded physical $K$-spin problem.

\begin{figure*}
\includegraphics[viewport=100bp 100bp 850bp 500bp,clip,scale=0.55]{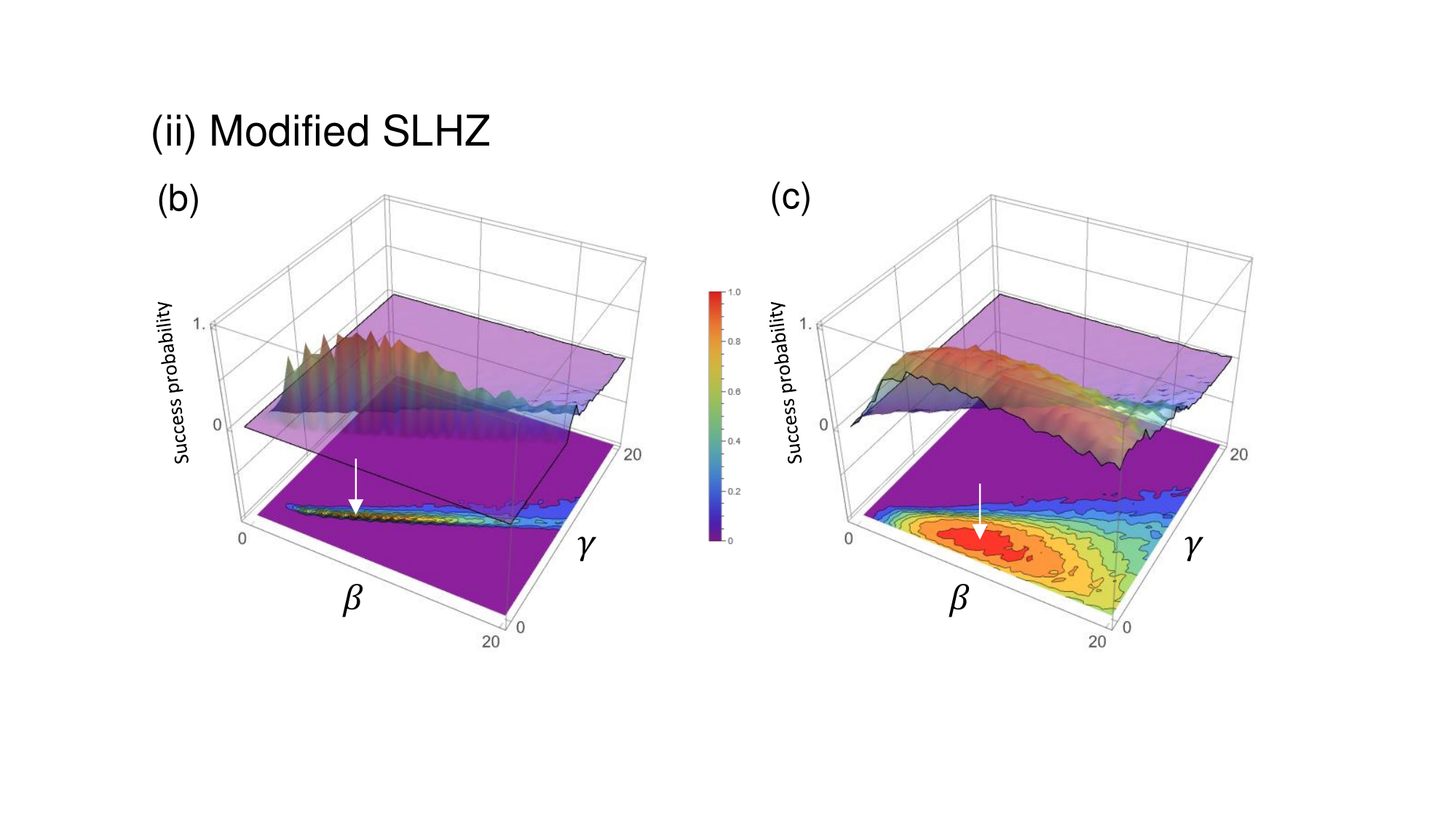}
\caption{Landscapes of the success probability distribution for the modified SLHZ scheme. The left and right plots are the results for (b) a classical annealing-based search and (c) a classical annealing-based search combined with postreadout BF decoding, respectively.}
\label{fig:4}
\end{figure*}

\begin{figure*}
\includegraphics[viewport=100bp 100bp 850bp 500bp,clip,scale=0.55]{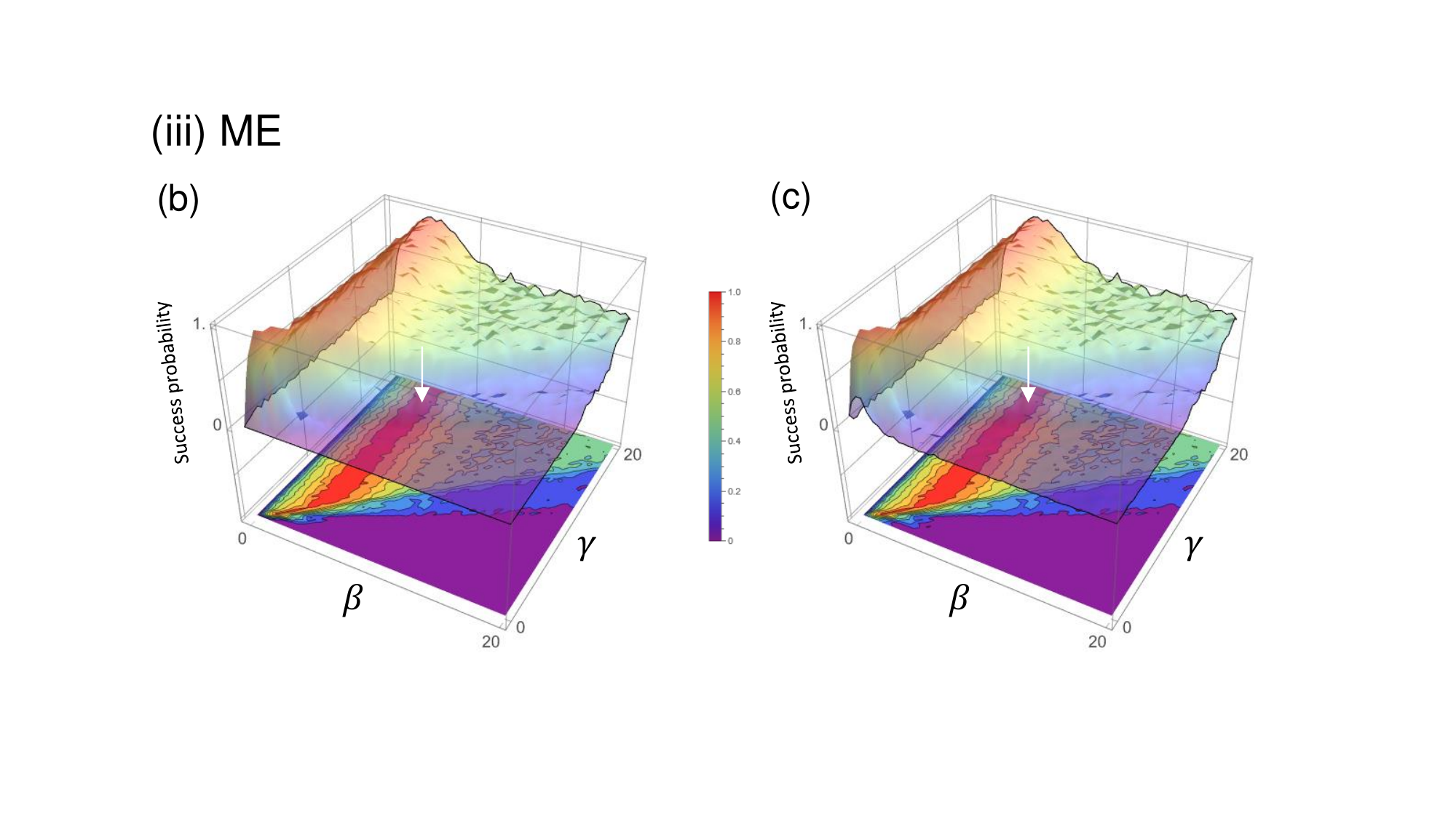}
\caption{Landscapes of the success probability distribution for the ME scheme. The left and right plots are the results for (b) a classical annealing-based search and (c) a classical annealing-based search combined with postreadout MV decoding, respectively.}
\label{fig:5}
\end{figure*}

To obtain $\textbf{z}^{*}$ efficiently, it is necessary to adjust the parameters $C_{i}$ $(i=4,3, \text{ or } ME)$ and the inverse temperature $\beta$ in the MCMC calculation. Since the Hamiltonian is given by Eqs. (\ref{eq:2}), (\ref{eq:5}), and (\ref{eq:9}), we adjusted $\beta$ and $\gamma=\beta C_{i}$ as independent parameters to be optimized. Figures \ref{fig:3}–\ref{fig:5} show the landscapes of the success probability distributions for the three embedding schemes (i) to (iii) associated with the experiments (b) and (c). These figures clearly demonstrate that, for the SLHZ and modified SLHZ schemes, the success probability distribution depends heavily on whether postreadout decoding is applied to the MCMC sampling readout. Furthermore, in the experiment (c), it was confirmed that $\textbf{z}^{*}$ can be obtained efficiently only in parameter regions $(\beta,\gamma)$ where the MCMC sample $\textbf{r}$ yields a non-code state that does not satisfy the parity constraints, regardless of the scheme used (see Ref. \cite{nambuPracticalHybridDecoding2026} for details). In contrast, in the ME scheme, the corresponding distribution is hardly affected by whether postreadout decoding is applied or not. This effectiveness of postreadout decoding is a key feature of the SLHZ method, which is based on LDPC codes.

\begin{figure*}
\includegraphics[viewport=70bp 0bp 850bp 600bp,clip,scale=0.65]{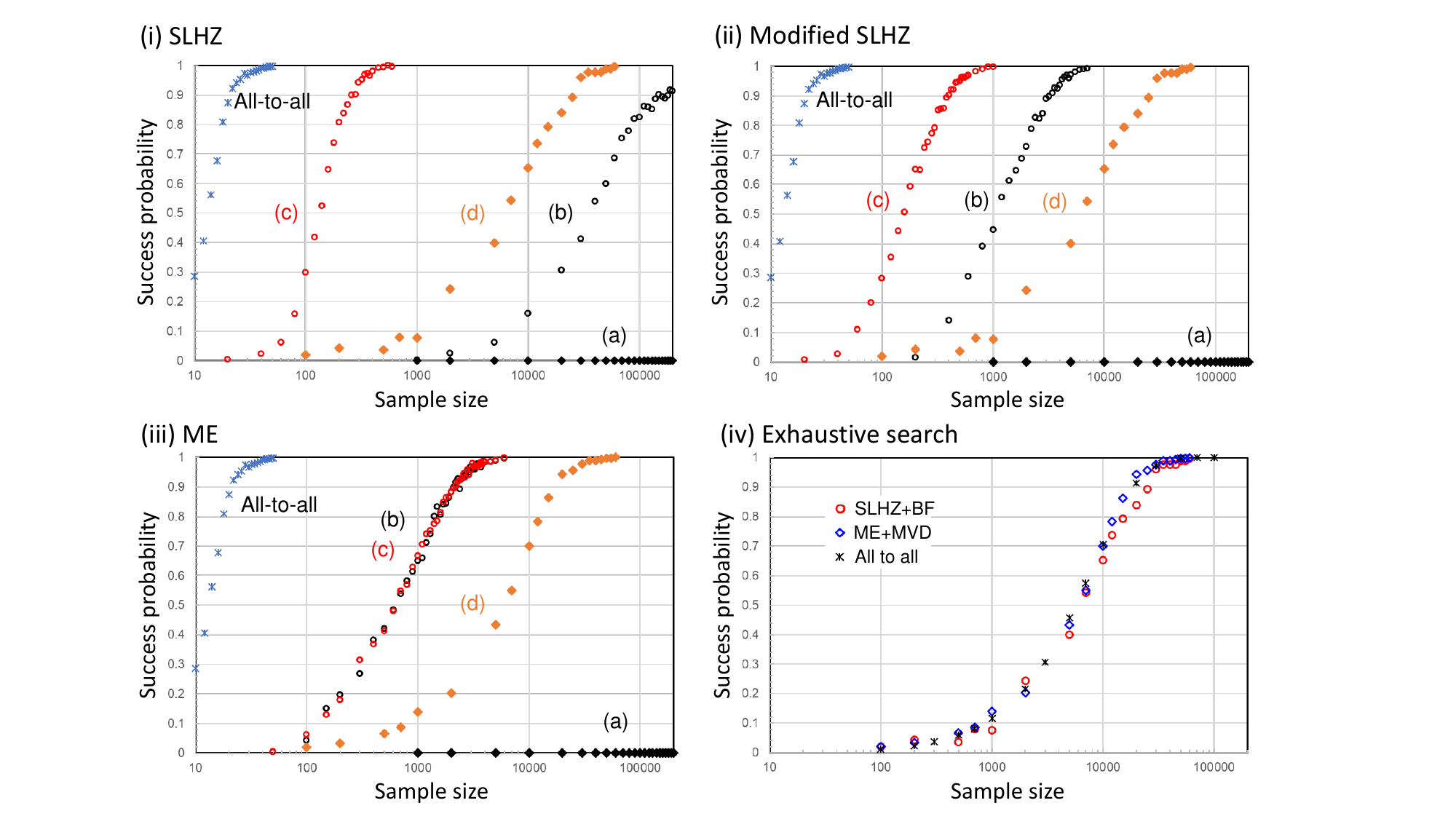}
\caption{Figures (i) to (iii): These show the relationship between the success probability and sample size for (i) the SLHZ scheme, (ii) the modified SLHZ scheme, and (iii) the ME scheme. Plots labeled by (a) to (d) correspond to the results of experiments (a) to (d) shown in Fig.\ref{fig:1}, respectively. Plots labeled “all-to-all” correspond to the result of annealing-based search using the logical Hamiltonian in Eq.\ref{eq:1}. Figure (iv): Relationship between success probability and sample size when combining exhaustive search with classical postreadout decoding in the SLHZ and ME schemes. For comparison, the result of an exhaustive search based on the all-to-all connected logical $N$-spin system is also plotted as “all-to-all”.}
\label{fig:6}
\end{figure*}

In this work, all the benchmarking experiments (c) and (d) were performed under the optimal conditions. The white arrows in Figs. \ref{eq:3}-\ref{eq:5} indicate the relevant optimal conditions for the adjustable parameters $(\beta,\gamma)$ for every experiment (c) and (d). Figures \ref{fig:6} (i)-(iii) show the success probability plotted against the sample size (in logarithmic scale) for every experiment (a)-(d) and every scheme evaluated under the optimal conditions. On the other hand, Fig.\ref{fig:6} (iv) shows the success probabilities concerning the experiment (d) for the SLHZ and ME schemes, as well as the exhaustive search based on the all-to-all connected logical $N$-spin system.  These figures imply the following insights:
\begin{figure*}
\includegraphics[viewport=100bp 100bp 850bp 500bp,clip,scale=0.55]{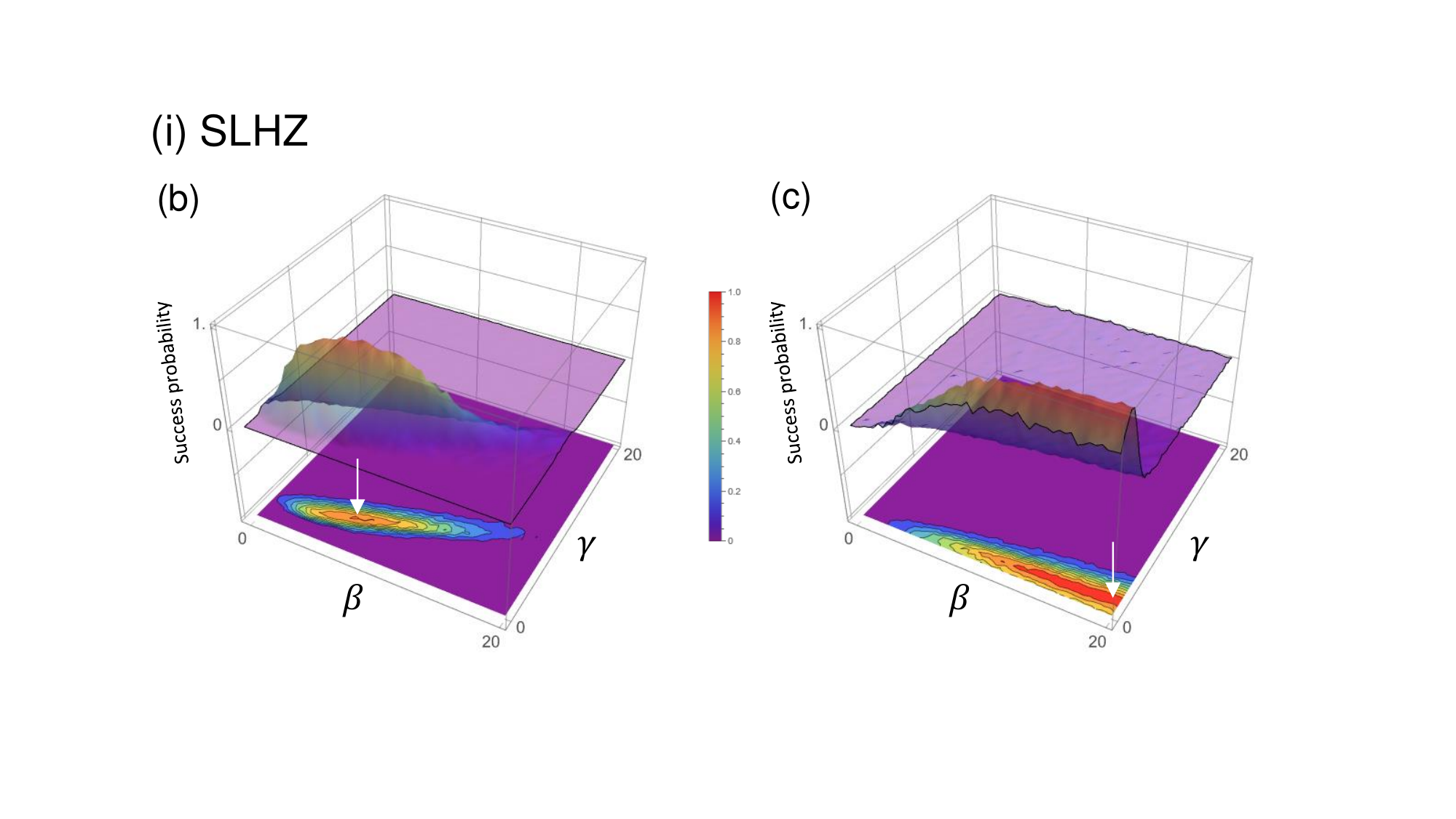}
\caption{Landscapes of the success probability distribution for the SLHZ scheme plotted as functions of the Lagrange weights $(\beta,\gamma=\beta C_{i})$. The left and right plots are the results for (b) a classical annealing-based search and (c) a classical annealing-based search combined with postreadout BF decoding, respectively. White arrows indicate optimal conditions for the weights $(\beta,\gamma)$.}
\label{fig:3}
\end{figure*}
\begin{enumerate}
    \item  The plots for the experiment (a) in Figs.  \ref{fig:6} (i)-(iii)  indicate that the success probability by the exhaustive search is negligible for all the schemes.
    \item Figure \ref{fig:6} (iv) compares the performance of the exhaustive search combined with classical postreadout decoding, i.e., the experiment (d), for the SLHZ and ME schemes. We find that they are comparable. Moreover, they are comparable to the performance of the exhaustive search based on the all-to-all connected logical $N$-spin system. 
    \item  In the SLHZ scheme shown in Fig. \ref{fig:6}  (i), the success probability for the classical annealing-based search, i.e., the experiment (b), is smaller than that for the exhaustive search combined with classical postreadout decoding, i.e., the experiment (d), whereas the relation is exchanged for the other two schemes as shown in Figs. \ref{fig:6} (ii) and (iii). As a result, as for classical annealing-based searches, the success probability of the SLHZ scheme is significantly lower than that of the modified-SLHZ and ME schemes.
    \item Comparing the SLHZ and modified SLHZ schemes, the success probability for the experiment (c) is larger than that for the experiment (b). This means that classical postreadout decoding enhances the performance of an annealing-based search for the SLHZ-based schemes. In particular, the enhancement is much greater for the SLHZ scheme than for the modified SLHZ scheme, because the performance of the experiment (b) is much worse for the SLHZ scheme than for the modified SLHZ scheme. 
    \item In the ME scheme shown in Fig. \ref{fig:6} (iii), the performance for the experiments (b) and (c) is comparable, i.e., the classical postreadout decoding does not enhance the performance of an annealing-based search. 
    \item The performance of the annealing-based searches combined with the classical postreadout decoding, i.e, the experiments (c), for the SLHZ and modified SLHZ schemes is comparable, as shown in Figs. \ref{fig:6} (i) and (ii). However, it is better than that of the ME scheme shown in Fig. \ref{fig:6}  (iii).
    \item The performance of the three embedding schemes never exceeds that of the classical annealing-based search based on the all-to-all connected logical Ising Hamiltonian (\ref{eq:1}).
\end{enumerate}

In what follows, we present brief discussions about the above results. First of all, although the number $K=\binom{N}{2}$ of spins is the same for the SLHZ and modified SLHZ schemes, the performance for the classical annealing-based search significantly differs between these schemes. This difference evidently comes from the different choice of the penalty Hamiltonian. We should note that the penalty Hamiltonian of the SLHZ scheme involves only short-range four-body interactions among the adjacent spins, while that of the modified SLHZ scheme involves many long-range three-body interactions among three spins. While the SLHZ scheme uses only short-range couplings, which allows device implementation with a quasi-planar layout, it inevitably sacrifices spin-updating characteristics because the effects of long-range couplings are reproduced through the cooperative action of numerous short-range couplings. As a result, the performance of an annealing-based search in the SLHZ scheme is quite unsatisfactory. However, this issue can be avoided by using classical decoding for the readout, which maps any physical state to its nearest code state, i.e., the physical state that satisfies the parity constraints. The function of this classical decoding is clearly shown in Fig.\ref{fig:6} (iv), where we present results obtained when a randomly chosen physical state was applied to the classical decoding. In this case, the decoded state should be the random code state associated with the random logical state. Then, the performance of the exhaustive search combined with classical postreadout decoding should be comparable to the performance of the exhaustive search for the original $N$-spin problem. This is actually confirmed in Fig.\ref{fig:6} (iv). It implies that we can regard the classical decoding as an efficient deterministic algorithm for finding the ground state of the penalty Hamiltonian, i.e., code state in the SLHZ and modified SLHZ schemes.

The additional point to note is that there is a problem with the penalty Hamiltonian of the SLHZ scheme that has been overlooked until now: it implicitly assumes that the weight $C_{4}$ remains constant across the possible set of four indices $\{i, j, k, l\}$ for the penalty term. This assumption is questionable from the viewpoint of the ECC. This is because not all information in the weight-4 syndromes defined in the SLHZ scheme is equally important. For example, there are syndromes whose indices involve the same index, i.e., $s_{ijjl}^{(4)}(\textbf{z})=z_{ij}z_{jj}z_{jl}z_{il}=z_{ij}z_{jl}z_{il}$, where $z_{jj}=1$ has been assumed to be fixed. Then this fixed variable cannot cause an error. In this case, since the error rate for this syndrome is lower than that for the other syndromes, the weights associated with it should be increased relative to those associated with the other syndromes. Therefore, the Hamiltonian (\ref{eq:2}) is not necessarily the adequate choice for use as an optimization cost function. In contrast, it is reasonable to choose the constant weight $C_{3}$ across the possible set of three indices $\{i, j, k\}$ for the penalty term in the Hamiltonian \ref{eq:5} , since the weight-3 syndromes treat all variables symmetrically, and they should be equally important.

Figures \ref{fig:6} (i) and (iii) imply that while the performance of the annealing-based search for the SLHZ scheme is significantly inferior to that for the ME scheme, which was pointed out in Ref. \cite{albashSimulatedquantumannealingComparisonAlltoall2016}, this relationship is reversed when postreadout classical decoding is combined with the annealing. Although it is obvious that this discussion neglects the overhead introduced by classical decoding, this overhead is much more insignificant than the cost of MCMC sampling, because our classical decoding is a parallel algorithm that runs in polynomial time. Therefore, although the SLHZ scheme alone introduces significant overhead, we conclude that by incorporating classical postreadout decoding, the SLHZ method has the potential to outperform the ME method.

To summarize, we have investigated and compared the performance of annealing-based search for the SLHZ and ME schemes, as well as related schemes, using a toy model. We have considered the advantages and disadvantages of the SLHZ scheme. Although the SLHZ scheme has the advantage of removing long-range couplings and enabling quasi-planar layout and implementation of the QA device using on-chip superconducting circuits, it inevitably incurs a large overhead. Nevertheless, this large overhead can be canceled by introducing the postreadout classical decoding. The postreadout BF decoding can efficiently find the code state defined in the SLHZ scheme, which allows us to improve the performance of annealing-based search for the optimal solution. Since QA devices based on the SLHZ method are still in development, this study employed classical MCMC sampling to simulate annealing-based searches. However, we believe this approach is sufficient to gain a deeper understanding of the core issue of significant overhead in the SLHZ method and to propose solutions. Our study revealed the possibility of the SLHZ scheme over the ME scheme as near-term QA devices. Nonetheless, further research will be required for future practical application, such as verifying the effectiveness of the hybrid decoding using actual QA devices. 

\begin{acknowledgments}
I would like to thank Dr. Masayuki Shirane of NEC Corporation and the National Institute of Advanced Industrial Science and Technology for his continuous support. This paper is partly based on results obtained from a project, Grant No. JPNP16007, commissioned by the New Energy and Industrial Technology Development Organization (NEDO), Japan. 
\end{acknowledgments}

\bibliography{APS}
\end{document}